\documentclass[aps, pre, twocolumn]{revtex4}

\usepackage{amsmath}
\usepackage{graphicx}
\usepackage[usenames]{color}
\usepackage{ifthen}
			
\definecolor{myred}{rgb}{0.8,0,0}
\definecolor{mygreen}{rgb}{0,0.6,0}
\definecolor{myblue}{rgb}{0,0,0.6}
\definecolor{myorange}{rgb}{1,0.55,0}
\definecolor{mygray}{rgb}{0.6,0.6,0.6}
\usepackage[
    pdfpagemode=none,
    pdfpagelabels=true,
    bookmarks=true,
    colorlinks,
    linkcolor=myblue,
    citecolor=mygreen,
    pagecolor=myred,
]{hyperref}
\usepackage{hyperref}
\usepackage{natbib}

\newcommand{\myFigRef}[1]{~\ref{f:#1}}
\newcommand{\myEqRef}[1]{~\ref{e:#1}}

\newcommand{\myEtAl}{{\em{~et~al.}}}

\newcommand{\mySub}[1]{_\mathrm{\scriptstyle #1}}

\newcommand{\Rg}{\mathit{R}_\mathrm{\scriptstyle g}}
\newcommand{\sO}{\mathit{s}_\mathrm{\scriptscriptstyle 0}}

\newcommand{\myRefs}[1]{
    \ifthenelse{\equal{#1}{}}
        {\textbf{\textcolor{myred}{\fontfamily{phv}\selectfont REFS: [...]}}}
        {\textbf{\textcolor{myred}{\fontfamily{phv}\selectfont REFS: [#1]}}}
}

\newlength{\figwidth}
\begin{document}

\title{Non-driven polymer translocation through a nanopore: \\computational evidence that the escape and relaxation processes are coupled}
\author{Michel G. Gauthier\footnote{E-mail: gauthier.michel@uOttawa.ca}}
\author{Gary W. Slater\footnote{E-mail: gary.slater@uOttawa.ca}}
\affiliation{Department of Physics, University of Ottawa, 150 Louis-Pasteur, Ottawa, Ontario K1N 6N5, Canada}
\date \today

\begin{abstract}
Most of the theoretical models describing the translocation of a polymer chain through a nanopore use the hypothesis that the polymer is always relaxed during the complete process. In other words, models generally assume that the characteristic relaxation time of the chain is small enough compared to the translocation time that non-equilibrium molecular conformations can be ignored. In this paper, we use Molecular Dynamics simulations to directly test this hypothesis by looking at the escape time of unbiased polymer chains starting with different initial conditions. We find that the translocation process is not quite in equilibrium for the systems studied, even though the translocation time $\tau$ is about $10$ times larger than the relaxation time $\tau_{\text{r}}$. Our most striking result is the observation that the last half of the chain escapes in less than $\sim12\%$ of the total escape time, which implies that there is a large acceleration of the chain at the end of its escape from the channel.
\end{abstract}

\maketitle

\newpage

\section{Introduction} \label{s:intro}
The translocation of polymers is the process during which a flexible chain moves through a narrow channel to go from one side of a membrane to the other. Many theoretical and numerical models of this fundamental problem have been developed during the past decade. These efforts are motivated in part by the fact that one of the most fundamental mechanism of life, the transfer of RNA or DNA molecules through nanoscopic biological channels, can be described in terms of polymer translocation models. Moreover, recent advances in manipulating and analyzing DNA moving through natural~\cite{Kasianowicz1996, Meller2001} or synthetic nanopores~\cite{Chen2004} strongly suggest that such mechanical systems could eventually lead to the development of new ultrafast sequencing techniques~\cite{Kasianowicz1996, Astier-Braha-Bayley, Deamer2002, Howorka-Cheley-Bayley, Kasianowicz-NatureMat, Lagerqvist-Zwolak-Ventra, Muthukumar2007, Vercoutere-Winters-Hilt-Olsen-Deamer-Haussler-Akeson,Wang-Branton}. However, even though a great number of theoretical~\cite{Sung1996, Muthukumar1999, Berezhkovskii2003, Flomenbom2003, Kumar2000, Lubensky1999, Ambjornsson2004, DiMarzio1997, Matsuyama2004, Metzler2003, Slonkina2003, Storm2005} and computational~\cite{Ali2005, Baumgartner1995, Chern2001, Chuang2001, Dubbeldam2007, Dubbeldam2007a, Farkas2003, Huopaniemi2006, Kantor2004, Kong2002, Loebl2003, Luo2006a, Luo2006, luo-2007-126, Matysiak2006, Milchev2004, Tian2003, Wei2007, Wolterink2006} studies have been published on the subject, there are still many unanswered questions concerning the fundamental physics behind such a process.

\begin{figure}[tb]
    \begin{center}
        \includegraphics[width=\figwidth]  {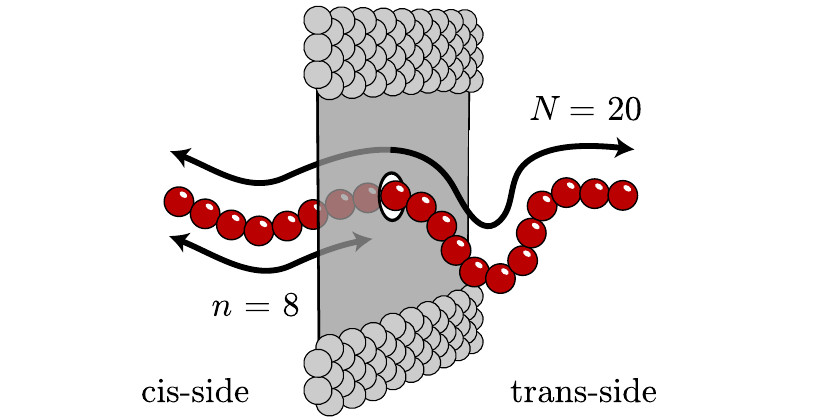}
        \caption{Schematic representation of our simulation system. The wall consists in a single layer of beads
                on an triangular lattice while the pore itself is formed by simply removing one wall bead (some wall beads
                and all of the solvents beads have been removed for clarity reasons).
                This simulation system is described in details in Ref~\cite{Guillouzic2006, GauthierMD2007}.
                The trans-side of the membrane is defined as the side where the chain terminates its translocation process (its final destination).
                The translocation coordinate $s$ is defined as the ratio of the number of monomers
                on the cis-side of the membrane, $n$, to the total number of monomers in the chain $N$ ($0\leq s=n/N \leq 1$).
        }
        \label{f:schema}
    \end{center}
\end{figure}

The best known theoretical approaches used to tackle this problem are the ones derived by Sung and Park~\cite{Sung1996}, and by Muthukumar~\cite{Muthukumar1999}. Both of these methods study the diffusion of the translocation coordinate $s$, which is defined as the fractional number of monomers on a given side of the channel (see Fig.\myFigRef{schema}). Sung and Park use a mean first passage time (MFPT) approach to study the diffusion of the translocation coordinate. Their method consists in representing the translocation process as the diffusion of the variable $s$ over a potential barrier that represents the entropic cost of bringing the chain halfway through the pore. The second approach, derived by Muthukumar, uses nucleation theory to describe the diffusion of the translocation coordinate. Several other groups have worked on these issues (see Refs.~\cite{Berezhkovskii2003, Kumar2000, Lubensky1999, Slonkina2003} for example), and many were inspired by Sung and Park's and/or by Muthukumar's work.  However, such models assume that the subchains on both sides of the membrane remain in equilibrium at all times; this is what we call the \emph{quasi-equilibrium hypothesis}. This assumption effectively allows one to study polymer translocation by representing the transport of the chain using a simple biased random-walk process~\cite{Flomenbom2003, GauthierMC2007a, GauthierMC2007b}.

In the case of driven translocation, simulations monitoring the radius of gyration of the subchains on both sides of the membrane have shown that the chains are not necessarily at equilibrium during the complete translocation process~\cite{Luo2006a, Tian2003}. However, as far as we know, no direct investigation of the quasi-equilibrium hypothesis has been carried out so far for \textit{unbiased} translocations, although it is commonly used to conduct theoretical studies. For example, the fundamental hypothesis behind the one-dimensional model of Chuang\myEtAl~\cite{Chuang2001} is that the translocation time is much larger than the relaxation time so that the polymer would have time to equilibrate for each new value of $s$. Chuang\myEtAl~found that the translocation time should scale like $N^{9/5}$ and $N^{11/5}$ with and without hydrodynamic interactions, respectively.  Their assumption is indirectly supported by the observation made by Guillouzic and Slater~\cite{Guillouzic2006} using Molecular Dynamics simulation with explicit solvent that the scaling exponent of the translocation time $\tau$ with respect to the polymer length $N$ ($\tau \sim N^{2.27}$) is larger than the one measured for the relaxation time ($\tau\mySub{r} \sim N^{1.71}$). We recently made similar observations for larger nanopore diameters~\cite{GauthierMD2007}. The main goal of the current paper is to carry out a \textit{direct test} of the fundamental assumption that is behind most of the theoretical models of translocation: that the chain can be assumed as relaxed at all times during the translocation process (the quasi-equilibrium hypothesis). We will be using two sets of simulations to compare the translocation dynamics of chains that start with the same initial value of $s$ but that differ in the way they reached this initial state.

\section{Simulation Method} \label{s:method}
We use the same simulation setup as in our previous publication~\cite{Guillouzic2006, GauthierMD2007}. In short, we use coarse-grained Molecular Dynamics (MD) simulations of unbiased polymer chains initially placed in the middle of a pore perforated in a one bead thick membrane (see Fig.\myFigRef{schema}). The simulation includes an explicit solvent. All particles interact via a truncated (repulsive part only) Lennard-Jones potential and all connected monomers interact via a FENE (Finitely Extensible Nonlinear Elastic) potential. The membrane beads are held in place on a triangular lattice using an harmonic potential and the pore consists in a single bead hole. All quantities presented in this paper are in standard MD units; i.e. that the lengths and the energies are in units of the characteristic parameters of the Lennard-Jones potential $\sigma$ and $\epsilon$, while the time scales are measured in units of $\sqrt{m\sigma^2/\epsilon}$ where $m$ represents the mass of the fluid particles. The simulation box size is of $\sim 28.1\sigma \times 29.2\sigma \times 27.5\sigma$, where the third dimension is the one perpendicular to the wall, and periodic boundary conditions are used in all directions during the simulation. We refer the reader to Ref.~\cite{Guillouzic2006, GauthierMD2007} for more details. Note that this simulation setup was shown to correctly reproduce Zimm relaxation time scalings~\cite{GauthierMD2007}.  

The simulation itself is divided into two steps; (1) the warm-up
period during which the $i^{\text{th}}$ bead of the polymer is kept
fixed in the middle of the pore while its two subchains are
relaxing on opposite sides of the wall, and (2) the translocation
(or escape) period itself during which the polymer is completely
free to move until all monomers are on the same side of the membrane
(note that the final location of the chain defines the
\textit{trans}-side of the membrane in this study since we have no
external driving force that would define a direction for the
translocation process). The time duration of the first period was
determined from previous simulations~\cite{Guillouzic2006, GauthierMD2007} using the
characteristic decay time of the autocorrelation function of the
chain end-to-end vector. The time elapsed during the second period
is what we refer to as the translocation time $\tau$.

In previous papers~\cite{Guillouzic2006, GauthierMD2007}, we
calculated both the relaxation time $\tau\mySub{r}(N)$ and the
translocation time $\tau (N)$ for polymers of lengths $N$ between 15
to 31 monomers in the presence of the same membrane-pore system. Our
simulation results, $\tau \approx 1.38N^{2.3}$ and $\tau\mySub{r}
\approx 0.43N^{1.8}$ in MD units, indicate that the escape time is
at least 10 times longer than the relaxation time for this range of
polymer sizes. These translocation times correspond to polymers
starting halfway through the channel and the relaxation times were
calculated with the center monomer (i.e. monomer $i= (N+1)/2$, where
$N$ is an odd number) kept fixed in the middle of the pore.

\begin{figure}[tb]
    \begin{center}
        \includegraphics[width=\figwidth]  {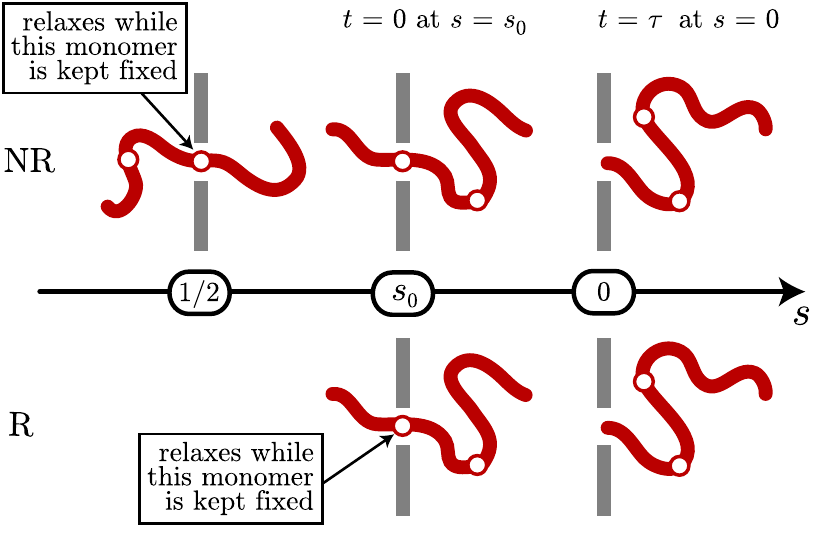}
        \caption{Schematic representation of our two sets of simulations, called R (for \textbf{R}elaxed) and NR (for \textbf{N}ot \textbf{R}elaxed).
                For the NR case, the middle monomer is kept fixed inside
                the pore during the initial warm-up relaxation
                phase. The polymer then moves freely until it completely escapes from the pore. However, the translocation clock then starts only
                when the polymer reaches state $s=\sO$ for the \textit{first} time. In the
                R case, the polymer is initially prepared in the $s=\sO$ state and allowed to relax with its
                $(N\sO+1)^{\textrm{th}}$ monomer fixed inside the pore. The translocation clock then starts immediately after the chain is
                released. The two sets of simulations thus differ
                only in the way the initial chain is prepared.
        }
        \label{f:2sims}
    \end{center}
\end{figure}

As we mentioned in the Introduction, the goal of this paper is to
run two different sets of simulations in order to \textit{directly}
test the quasi-equilibrium hypothesis (see Fig.\myFigRef{2sims}). In
the first type of simulations (that we will call NR for \emph{Not
Relaxed}), we start with the same configuration as in the previous
paper: the polymer chain is initially placed halfway through the
pore, then allowed to relax with its middle monomer fixed, and is
finally released. However, we do not start to calculate the
translocation time from that moment; instead, we wait until the
translocation coordinate has reached a particular value $s=\sO$ for
the first time (see the top part of Fig.\myFigRef{2sims}). The
translocation time $\tau^{\mathrm{NR}}(\sO)$ thus corresponds to a
chain that starts in state $s=\sO$ with a conformation that is
affected by the translocation process that took place between states
$s=1/2$ and $s=\sO$. In the second series of simulations (called R
for \emph{Relaxed}), we allow the chain to relax in state $s=\sO$
before it is released. In other words, the $(N\sO+1)^{\textrm{th}}$
monomer is fixed during the warm-up period (see the bottom part of
Fig.\myFigRef{2sims}); the corresponding translocation time
$\tau^{\mathrm{R}}(\sO)$ now corresponds to a chain that is fully
relaxed in its initial state $s=\sO$. Obviously, the
quasi-equilibrium hypothesis implies the equality
$\tau^{\mathrm{NR}}(\sO)=\tau^{\mathrm{R}}(\sO)$, a relationship
that we will be testing using extensive Molecular Dynamics
simulations. In both cases, we include all translocation events in
the calculations, including those that correspond to backward
translocations (i.e., translocations towards the side where the
smallest subchain was originally found).

\section{NR vs R: The escape times}
\begin{figure}[tb]
    \begin{center}
        \includegraphics[width=\figwidth]  {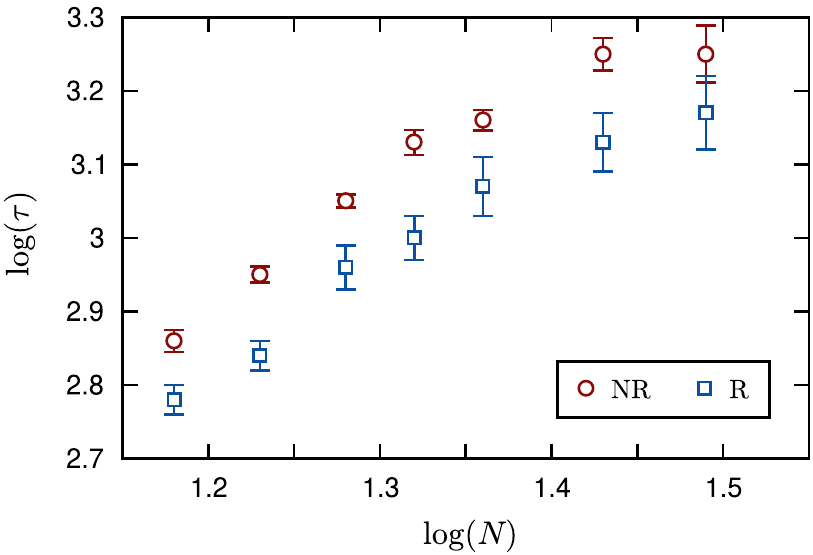}
        \caption{
            Translocation times $\tau$ for relaxed (R) and not relaxed (NR) polymers. 
            The initial condition is $\sO=6/N$ for all molecular
            sizes. 
        }
        \label{f:trans_centered6}
    \end{center}
\end{figure}
Figure\myFigRef{trans_centered6} shows the translocation times
obtained from these two sets of simulations when we choose the
starting point $\sO=6/N$ (six monomers on one side of the wall, and
all the others on the other side). We clearly see that the
translocation process is faster when the polymer is initially
relaxed (R). The difference between the two escape times is around
$25\%$ for all polymer lengths $N$. Since the relaxation state of
the chain at $s=\sO$ is the only difference between the two set of
results, this indicates that the NR polymers are not fully relaxed
at $s=\sO$. Thus, contrary to the commonly used assumption, even an unbiased polymer is not in quasi-equilibrium during its translocation process.

Also interesting is the probability to escape on the side where the
longest subchain was at the beginning of the simulation. We observed
(data not shown) that this probability was always $\sim 10-20\%$
times larger in the R simulations. This observation also confirms
the fact that the chain is out of equilibrium during translocation
since its previous trajectory even affects the final outcome of the
escape process. Note that we did verify that this difference is not
the reason why the escape times are different.

\section{NR vs R: The radii of gyration}
As we will now show, the slower NR translocation process is due to a non-equilibrium compression of the subchain located on trans-side of the wall. By compression, we mean that the radius of gyration $\Rg$ of that part of the polymer is smaller than the one it would have if it were in a fully relaxed state.

\begin{figure}[tb]
    \begin{center}
        \includegraphics[width=\figwidth]  {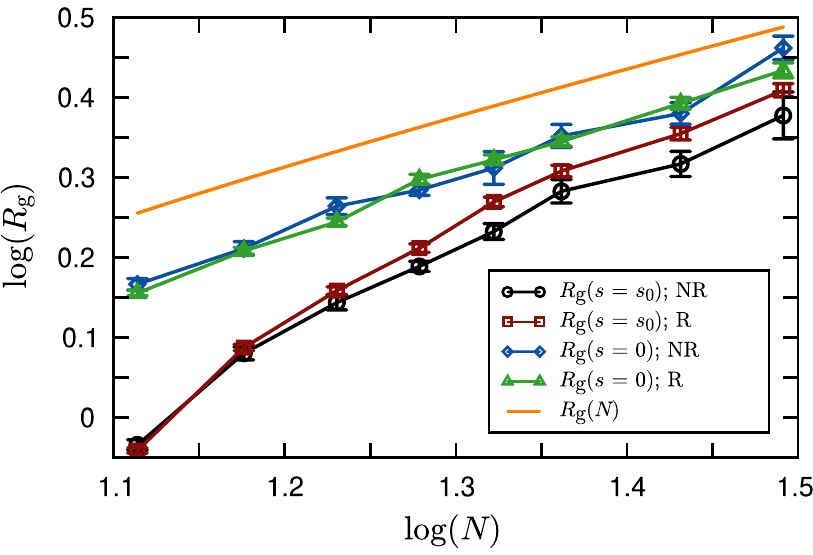}
        \caption{
           	The radius of gyration of the longest subchain vs. polymer size $N$.
            We show values corresponding to the beginning ($s=\sO=6/N$) and the end ($\sO=0$) of the
            process, both for chains that were initially relaxed (R) and non-relaxed (NR).
            The fifth data set is the radius of gyration of a relaxed chain of length $N$~\cite{Guillouzic2006, GauthierMD2007}.
        }
        \label{f:rg}
    \end{center}
\end{figure}

Figure\myFigRef{rg} compares the mean radius of gyration of the subchain on the trans-side at $s=\sO$ for both the relaxed (R) and the non-relaxed (NR) states (the two first curves from bottom). The radius of gyration is larger for the relaxed state when the number of monomers is greater than about 19, i.e. $\Rg^{\mathrm{R}}(s=\sO) > \Rg^{\mathrm{NR}}(s=\sO)$ if $N>19$. This is the second result that suggest the translocation process is not close to equilibrium. Moreover, this discrepancy between the two states increases with $N$ (the two curves diverge) over the range of polymer lengths studied here. Figure\myFigRef{rg} also shows that this difference is negligible by the time the escape is completed ($\Rg^{\mathrm{R}}(s=0) \approx \Rg^{\mathrm{NR}}(s=0)$). However, it is important to note that the final radius of gyration is always smaller than the value we would obtain for a completely relaxed chain of size $N$ (the top line, $\Rg\approx 0.357\,N^{0.631}$). Of course, this means that the R simulations, which start with equilibrium conformations, also finish with non-equilibrium states.

\section{The \lowercase{\textsl{s}(\textsl{t})} curve}
\begin{figure}[tb]
    \begin{center}
        \includegraphics[width=\figwidth]  {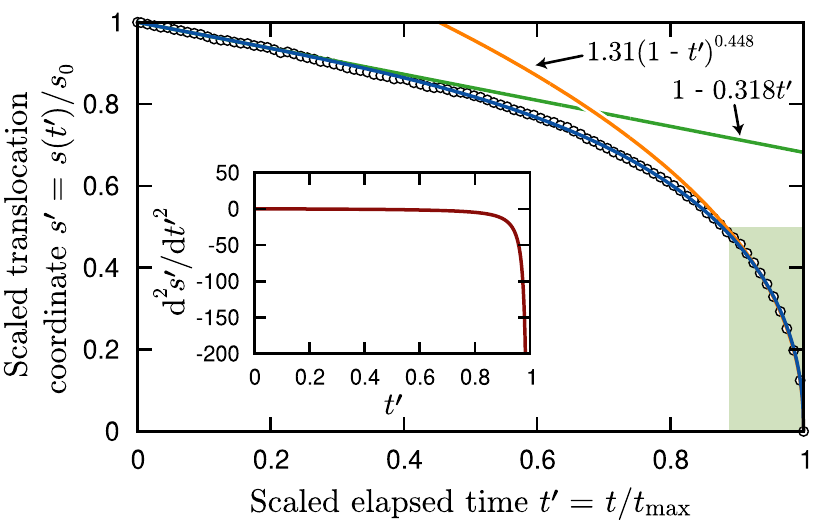}
        \caption{
            Scaled translocation coordinate $s^\prime=s(t^\prime)/\sO$ as a function of scaled time $t^\prime=t/t\mySub{max}$,
            where $t\mySub{max}$ is the individual translocation time for each translocation event that was simulated.
            Eight curves (not shown) were obtained for $N=13,\,15,\,17,\,19,\,21,\,23,\,27,\,\mathrm{and}\; 31$ the following way:
            for a given chain length initially placed halfway through the pore,
            (1) each of the translocation events gives a $s(t)$ curve that goes from $s(0)=1/2$ to $s(t\mySub{max})=0$,
            (2) then each of these curves is rescaled in time using $t^\prime=t/t\mySub{max}$,
            (3) and finally, the time-axis is discretized and all the curves for that given $N$ are averaged along the $y$-axis.
            Data points (circles) are the average of these eight curves which are not shown since their distribution was of the order of the data point sizes.
            The solid line that fits the universal curve represented by the complete data set is given by Eq.\myEqRef{s_vs_t}.
            The inset presents the acceleration of the scaled translocation coordinate
            $\mathrm{d}^2 s^{\prime}/\mathrm{d} {t^\prime}^2$ obtained from Eq.\myEqRef{s_vs_t}.
        }
        \label{f:s_vs_t}
    \end{center}
\end{figure}

Why do we observe such a large amount of compression when the translocation time is more than ten times larger than the relaxation time? A factor of ten would normally suggest that a quasi-equilibrium hypothesis would be adequate. The answer to this question is clearly illustrated in Fig.\myFigRef{s_vs_t} where we look at the normalized translocation coordinate $s^\prime=s(t^\prime)/\sO$ as a function of the scaled time $t^\prime$. These NR simulations used the initial condition $\sO=1/2$ (thus starting with symmetric conformations and maximizing the escape times). For a given polymer length, each $s(t)$ curve (we have typically used $\sim500$ runs per polymer length) was rescaled using its own escape time $t\mySub{max}$, such that $t^\prime=t/t\mySub{max}$ and $0 \leq t^\prime \leq 1$. These curves were then averaged to obtain eight rescaled data sets (one for each molecular size in the range $13\leq N \leq 31$; note that $N$ must be an odd number). Remarkably, the eight rescaled curves were essentially undistinguishable (data not shown). This result thus suggests that the time evolution of the translocation coordinate $s(t)$ follows a \textsl{universal} curve; the latter, defined as an average over all molecular sizes, is shown (circles) in Fig.\myFigRef{s_vs_t}. Please note that the translocation coordinate is defined with respect to the final destination of the chain ($s=N\mySub{cis}/N$), and not the side with the shortest subchain at a given time.


This unexpected universal curve has two well-defined asymptotic
behaviors: (1) for short times, we observe the apparent linear functional
\begin{equation}
    s^\prime (t^\prime)= 1-0.318 t^\prime \,,
\end{equation}
which we obtain using only the first 10\% of the data, and (2) as
$t^\prime \rightarrow 1$, the average curve decays rapidly towards
zero following the power law relation
\begin{equation}
    s^\prime (t^\prime)= 1.31 \times (1-t^\prime)^{0.448} \,,
\end{equation}
this time using the last 10\% of the data. The whole data set can
then be fitted using the interpolation formula
\begin{equation}
    s^\prime (t^\prime)= (1+0.130\,t^\prime+0.216\,{t^\prime}^2)\times(1-t^\prime)^{0.448} \,,
    \label{e:s_vs_t}
\end{equation}
where the coefficient of the ${t^\prime}^2$ term is the only
remaining fitting parameter. Equation\myEqRef{s_vs_t} is the solid
line that fits the complete data set in Fig.\myFigRef{s_vs_t}. As we
can see, this empirical fitting formula provides an excellent fit.

Figure\myFigRef{s_vs_t} can be viewed as the percentage of the
translocation process (in terms of the number of monomers that have
yet to cross the membrane in the direction of the trans side) as a
function of the percentage of the (final translocation) time elapsed
since the beginning. The small shaded region in
Fig.\myFigRef{s_vs_t} represents the second \textit{material} half
(as opposed to \textit{temporal} half) of the escape process
($s=\sO/2$). However, this region approximatively covers only the
last $\sim 12\%$ of the rescaled time axis; this clearly implies a
strong acceleration of the chain at the end of its exit. The first
$50\%$ of the monomer translocations take the first $\sim 88\%$ of
the total translocation time. The inset in Fig.\myFigRef{s_vs_t}
emphasizes the fact that the translocation coordinate is submitted
to a strong acceleration at the late stage of the translocation
process.

This large acceleration of the translocation process is
entropy-driven. At short times, the difference in size between the
two subchains is small, and entropy is but a minor player. At the
end of the process, however, this difference is very large and the
corresponding gradient in conformational entropy drives the process,
thus leading to a positive feedback mechanism. Translocation is then
so fast that the subchains cannot relax fast enough and the
quasi-equilibrium hypothesis fails. The trans-subchain is compressed
because the monomers arrive faster than the rate at which this coil
can expand. The ratio of ten between the translocation time and the
relaxation time (for the polymer lengths and initial conditions that
we have used) is too small because half of the translocation takes
place in the last tenth of the event.

Finally, the existence of a universal curve is a most interesting
result. Clearly, our choice of rescaled variables has allowed us to
find the fundamental mechanisms common to all translocation events.
This universal curve is expected to be valid as long as the radius
of gyration of the polymer chain is much larger than the pore size,
and it demonstrates that our results are not due to finite size
effects. Finally, we present in Appendix an asymptotic derivation to explain the apparent short time linear scaling of the translocation coordinate. This demonstration is based on the the fact that, in this particular limit, the motion is purely described by unbiased diffusion, a case for which we can do analytical calculations.

\section{The \textsl{R}\ensuremath{_\textrm{\textbf{g}}}(\textsl{t}) curve}
\begin{figure}[tb]
    \begin{center}
        \includegraphics[width=\figwidth]  {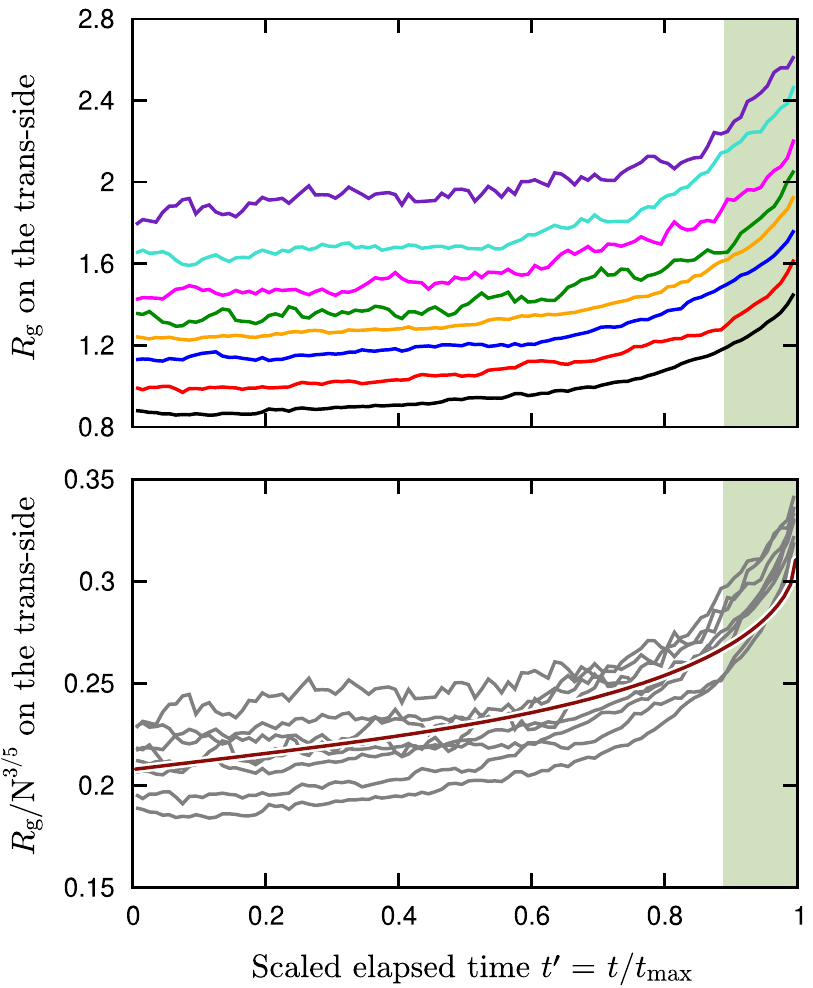}
        \caption{
                (a) Radius of gyration on the trans-side of the wall as a function of the scaled translocation time. Each simulation event is rescaled using the time $t\mySub{max}$ it took to exit the channel ($t^\prime=t/t\mySub{max}$). The scaled time is then always bounded between $0 \leq t^\prime \leq 1$. From bottom to top, the eight curves were obtained for $N=13,\,15,\,17,\,19,\,21,\,23,\,27,\,\mathrm{and}\;31$ by averaging $\Rg(t^\prime)$ over hundreds of simulations (typically $\sim500$ runs).
                (b) Rescaling of the curves presented in part (a). Each radius of gyration curve was divided by $N^{3/5}$ to obtain the gray curves ($\Rg^\prime = \Rg/N^{3/5}$).
                The smooth curve is given by Eq.\myEqRef{Rg_vs_s} with a proportionality constant of $0.315$.
                The shaded region covers the last $11\%$ of the translocation time and begins at the mid-point of the average radius of gyration increases, i.e. at ${\Rg^\prime}(t^\prime)\approx({\Rg^\prime}(1)+{\Rg^\prime}(0))/2$.
        }
        \label{f:Rg_vs_t}
    \end{center}
\end{figure}

Still more evidence that un-driven (no external field)
translocation is not a quasi-equilibrium process is presented in
Fig.\myFigRef{Rg_vs_t}a where we show how the mean radius of
gyration of the subchain located on the trans-side of the wall
changes with (rescaled) time during the NR translocation process
(like in the previous section, we have chosen the initial condition
$\sO=1/2$ here). All the curves have approximatively the same shape,
i.e. an initial period during which the radius of gyration increases
rather slowly, followed by an acceleration period that becomes very
steep at the end. When these curves are rescaled by a
three-dimensional Flory factor of $N^{3/5}$ ($\Rg^\prime(t^\prime)
\equiv \Rg(t^\prime)/N^{3/5}$, see Fig.\myFigRef{Rg_vs_t}b), they
seem to all fall approximatively onto each other. As we observed for
the translocation coordinate $(s^\prime)$, the radius of gyration
$\Rg(t^\prime)$ is experiencing a noticeable acceleration at the end
of the translocation process. Again, the shaded zone in
Fig.\myFigRef{Rg_vs_t} shows that the second half of the process
occurs in the last $\sim 11\%$ of the translocation time.

If we assume that Flory's argument ($\Rg \sim N^{3/5}$) is valid during the complete translocation process, we must be able to \textit{translate} the expression given by Eq.\myEqRef{s_vs_t} in order to fit the increase of the radius of gyration presented in Fig.\myFigRef{Rg_vs_t}b, i.e. we should have 
\begin{equation}
    \label{e:Rg_vs_s}
    \Rg^\prime (t^\prime) = b \times \left( 1 - \frac{s^\prime (t^\prime)}{2} \right)^{3/5} \,,
\end{equation}
where the $1 - s^\prime (t^\prime)/2$ represents the fraction of the chain that is on the trans-side at the time $t^\prime$ and $b$ is a length scale proportional to the Kuhn length of the chain. We used Eqs.\myEqRef{s_vs_t} and \myEqRef{Rg_vs_s} to fit the average of the eight $\Rg^\prime (t^\prime)$ curves presented in Fig.\myFigRef{Rg_vs_t}b and obtained $b=0.315$ (see the smooth curve). This one-parameter fit does a decent job until we reach about $80\%$ of the maximum time. However, it clearly underpredicts $\Rg$ in the last stage of the translocation process, i.e. during the phase of strong acceleration discussed previously. This observation also validates the fact that the translocation process is out of equilibrium during that period. In fact, the failure of the three-dimensional Flory's argument is also highlighted by the scaling of the radius of gyration at the end of the translocation process. Indeed, the third and fourth data sets presented in Fig.\myFigRef{rg} have a slope that is around $0.73$, which is closer to the two-dimensional Flory's scaling of $\Rg \sim N^{3/4}$.

\section{Conclusion}
In summary, we presented three different numerical results that contradict the hypothesis that polymer translocation is a quasi-equilibrium process in the case of unbiased polymer chains in the presence of hydrodynamic interactions. First, we reported a difference in translocation times that depends on the way the chain conformation is prepared, with relaxed chains translocating faster than chains that were in the process of translocating in the recent past. Second, we saw that the lack of relaxation also leads to conformational differences (as measured by the radius-of-gyration $\Rg$) between our two sets of simulations; in fact, translocating chains are highly compressed. Third, perhaps the strongest evidence is the presence of a large acceleration of both the translocation process (as measured by the translocation parameter $s$) and the growth of the radius of gyration: roughly half of the escape actually occurs during a time duration comparable to the relaxation time! The large difference between the mean relaxation and translocation times is not enough to insure the validity of the quasi-equilibrium hypothesis under such an extreme situation. It is important to note, however, that a longer channel would increase the frictional effects (and hence the translocation times) while reducing the entropic forces on both sides of the wall; we thus expect the quasi-equilibrium hypothesis to be a better approximation in such cases.

The curve presented in Fig.\myFigRef{s_vs_t} is quite interesting.
It demonstrates that the translocation dynamic is a highly nonlinear
function of time. We proposed an empirical formula
(Eq.\myEqRef{s_vs_t}) to express the evolution of the translocation
coordinate as a function of time (both in rescaled units) that
provides an excellent fit to our simulation data. Based on Flory's
argument for a three-dimensional chain, we presented a second
expression (Eq.\myEqRef{Rg_vs_s}) of a similar form for the increase
of the radius of gyration during the translocation process. However,
this relationship is not valid for the complete translocation
process, yet more evidence of the lack of equilibrium at the late
stage of the chain escape.

Finally, going back to the question in the title of this article, we conclude that the chain shows some clear signs of not being in a quasi-equilibrium state during unforced translocation (especially at the end of the escape process). However, although the difference is as large as $25\%$ when we start with only 6 monomers on one side, we previously demonstrated~\cite{Guillouzic2006, GauthierMD2007} that this simulation setup gives the expected scaling laws. The latter observation is quite surprising and leads to a non-trivial question: why scaling laws that were derived using a quasi-equilibrium hypothesis predict the proper dynamical exponents for chains that are clearly out of equilibrium during a non-negligeable portion of their escape? Perhaps the impact of these non-equilibrium conformations during translocation would be larger for thicker walls or stiffer chains; this remains to be explored. Obviously, the presence of an external driving force, such as an electric field, would lead the system further away from equilibrium; we thus speculate that there is a critical field below which the quasi-equilibrium hypothesis remains approximately valid, but beyond which the current theoretical exponents may have to be revisited.

This work was supported by a Discovery Grant from the Natural Sciences and Engineering Research Council of Canada (\emph{NSERC})
to GWS and by scholarships from the \emph{NSERC} and the University of Ottawa to MGG. The results presented in this paper were obtained using the computational resources of the SHARCNET and HPCVL networks.

\appendix*
\section{Derivation of the scaled translocation coordinat in the short time limit}
\label{s:a1}
In the short time limit, the translocation variable diffuses normally from its initial position (the potential landscape is very flat, see Fig.~2 in Ref.~\cite{Muthukumar1999}). In such a case, entropic pulling can be neglected and the translocation problem is equivalent to a non biased first-passage problem in which the displacement $x(t)$ from the initial position as a function of time grows following a Gaussian distribution. Consequently, if diffusion is normal, the scaled translocation coordinate is given by
\begin{equation}
s^{\prime}(t) = \int_{-\infty}^{\infty} \frac{1}{\sqrt{2Dt\pi}} \exp \left ( \frac{-x^2}{2Dt} \right ) \;   s^{\prime}(x)  \; dx \,,
\end{equation}
where D is the diffusion coefficient and $s^\prime(x) $ is the scaled translocation coordinate of the chain when it has moved over a curvilinear distance $x(t)$. According to our definition of the scaled translocation coordinate $s^\prime$, the latter value is given by
\begin{eqnarray}
s^\prime(x) &=& \frac{1}{s_0} \left \{ 
		\overbrace{\left ( \frac{1}{2} + \frac{\left | x \right | }{L} \right )}^{\mathrm{prob. \, exit \, same \, side}} 
		\underbrace{\left ( \frac{1}{2} - \frac{\left | x \right | }{L}  \right )}_{s(t)}  \right . \;\;\;\;\;\;\;\;\;\;\;\;\; \nonumber \\
		&& \left . 
		\;\;\;\;\;\;\; +\overbrace{\left ( \frac{1}{2} - \frac{\left | x \right | }{L}  \right )}^{\mathrm{prob. \, exit \, other \, side}} 
		\underbrace{\left ( \frac{1}{2} + \frac{\left | x \right | }{L}  \right )}_{s(t)} 
		\right \} \,,
\end{eqnarray}
where $L$ is the total length of the chain. The first term is the probability to exit on the side where the chain is, times the corresponding translocation coordinate ($s<0.5$). The second term refers to chains that will eventually exit on the other side of the channel ($s>0.5$). Remember that we defined the translocation coordinate with respect to the side where the chain eventually exits the channel. Consequently, the probabilities used in the last equation are obtained from the solution of the one-dimensional first-passage problem of  an unbiased random walker diffusing between two absorbing boundaries. The solution to this problem is explained in great details in Ref.~\cite{Redner2001}; the only result of interest for us is that the probabilities to be absorbed by the two boundaries are given by $0.5 \pm x^\prime$, where $x^\prime$ is the fractional distance between the particle position and the midpoint between the two boundaries. 

Combining the last two equations gives (using $s_0=1/2$)
\begin{eqnarray}
s^{\prime}(t) &=&  \int_{-\infty}^{\infty} \frac{1}{\sqrt{2Dt\pi}}\exp \left ( \frac{-x^2}{2Dt} \right ) \; \frac{1-4x^2/L^2}{2s_0} \; dx  \nonumber \\ 
&=&   \frac{1}{2s_0} \; \int_{-\infty}^{\infty} \frac{1}{\sqrt{2Dt\pi}}\exp \left (\frac{-x^2}{2Dt} \right ) \;  dx \nonumber \\
& &	
	+  \frac{2}{s_0} \int_{-\infty}^{\infty} \frac{1}{\sqrt{2Dt\pi}}\exp \left ( \frac{-x^2}{2Dt} \right ) \; \frac{x^2}{L^2} \; dx \nonumber \\
	&=& 1 - \frac{4Dt}{L^2} \,,
\end{eqnarray}
which predicts a linear decrease of the scaled translocation coordinate in the limit of very short time. One should note that the later derivation is strictly for short times since entropic pulling will eventually bias the chain translocation process. 

Finally, we can test our derivation by comparing our result with the linear regression presented in Fig.~5. This gives us that
\begin{equation}
\frac{0.318}{t_{\mathrm{max}}} = \frac{4D}{L^2} \,,
\end{equation}
which is equivalent to
\begin{equation}
2Dt_{\mathrm{max}} = (0.399L)^2\,.
\end{equation}
This means that, if entropic effects are neglected, a chain would travel a distance approximatively equal to 0.4 of its total length during a time equal to the observed translocation time. The fact that this result is smaller than $0.5L$ (the value corresponding to complete translocation) is an indication that entropy accelerates the escape of the chain. The slope of $-0.318$ indicates that translocation would be approximatively 3 times slower if entropic effects were cancelled. Finally, one should bear in mind that our linear decrease prediction is for normal diffusion only. However, Chuang~\myEtAl~predicted an anomalous diffusion exponent of 0.92~\cite{Chuang2001}. It would not be possible to observe the effect of such slightly subdiffusive regime with the precision of our data here.


\end{document}